# Magnetic Properties of a New Cobalt Hydrogen Vanadate with Dumortierite-Like Structure: $Co_{13.5}(OH)_6(H_{0.5}VO_{3.5})_2(VO_4)_6$

**Mohammed Hadouchi,[a]* Abderrazzak Assani,[a] Mohamed Saadi,[a] Abdelilah Lahmar,[b] Mimoun El Marssi[b] and Lahcen El Ammari[a]**

*[a]Laboratoire de Chimie Appliquée des Matériaux, Centre des Sciences des Matériaux, Faculty of Sciences, Mohammed V University in Rabat, Avenue Ibn Battouta, BP. 1014, Rabat, Morocco, and [b]Laboratoire de Physique de La Matière Condensée (LPMC), Université de Picardie Jules Verne, Amiens, France*

*Correspondence email: hadouchimohammed@yahoo.com*

## Abstract

The magnetic properties of a novel cobalt based hydrogen vanadate, $Co_{13.5}(OH)_6(H_{0.5}VO_{3.5})_2(VO_4)_6$, are reported. This new magnetic material was synthesized in single crystals form using a conventional hydrothermal method. Its crystal structure was determined from single crystal X-ray diffraction data and was also characterized by scanning electron microscopy. Its crystal framework belongs to dumortierite-like structure which consists of large hexagonal channels and triangular ones; the large hexagonal channels contains 1D chains of face-sharing $CoO_6$ octahedra linked to the framework by rings of $VO_4$ tetrahedra, while the triangular channels are occupied by chains of $V_2O_5$ pyramidal groups. The magnetic properties of this material were investigated by DC magnetic measurements and it shows the occurrence of antiferromagnetic interactions.

## 1. Introduction

Magnetic materials with geometrical frustration (Ramirez, 1994; Moessner & Ramirez, 2006) have attracted great attention due to the degeneracy of the spin ground state in these materials leading to fascinating magnetic properties, e.g., spin glass (Wang *et al.*, 2006) and spin liquid (Balents, 2010; Okamoto *et al.*, 2007). Moreover, the synthesis of new geometrically frustrated magnetic structures is a crucial challenge for chemists. Geometrical frustration in magnetic materials occurs from the incompatibility of nearest neighbor spin-spin interactions, when the system is antiferromagnetic, the antiparallel spin alignment with their neighbors remains frustrated, and therefore, the material develop a weak ferromagnetic state. This remarkable magnetic behavior is commonly observed in frustrated lattices such as Kagomé lattice (Moulton *et al.*, 2002), Pyrochlore (Harris *et al.*, 1997; Bramwell, 2001) , Star lattice (Zheng *et al.*, 2007) and Maple-Leaf lattice (Aliev *et al.*, 2012).

In this context, cobalt-based materials have attracted considerable attention due to their interesting magnetic properties such as geometrical frustration (Hardy *et al.*, 2003; Moessner & Ramirez, 2006) and magnetic anisotropy (Gambardella, 2003). Among cobalt-based materials that have been magnetically studied, e.g., $Ca_3Co_2O_6$ (Aasland *et al.*, 1997; Kageyama *et al.*, 1997), which was reported showing a field-induced transition from ferrimagnetic to ferromagnetic structure and strong magnetic anisotropy. Furthermore, a spin-glass-like dynamics were observed in the cobalt-based hydroxysulfate, $Na_2Co_3(OH)_2(SO_4)_3(H_2O)_4$ (Zhang *et al.*, 2008) with ferrimagnetic chains, whereas the authors (Vilminot *et al.*, 2010), showed a large coercivity of 70 kOe at 1.8 K in the ferrimagnet, $K_2Co_3(OH)_2(SO_4)_3(H_2O)_2$.

In the framework of exploring novel cobalt-based materials with new structural features, we report here for the first time the synthesis via hydrothermal method of a new cobalt hydrogen vanadate $Co_{13.5}(OH)_6(H_{0.5}VO_{3.5})_2(VO_4)_6$, with dumortierite-like structure and its magnetic properties.

## 2. Experimental

The morphology and semi-qu-anti-tative analysis of the synthesized single crystals were performed by JEOL JSM-IT100 InTouchScope™ scanning electron microscope (SEM) equipped with energy dispersive X-ray spectroscopy analyzer (EDS).

Single crystal with appropriate size was selected for the X-ray data collection at room temperature on a Bruker D8 VENTURE diffractometer equipped with a PHOTON II CPAD detector and IμS 3.0 microfocus X-ray source (Mo Kα radiation, $\lambda$ = 0.71073 Å). The software APEX3 was used for data collection and SAINT for cell refinement and data reduction. A total number of 32344 reflections were measured in the range of $\theta_{min}$ = 3.2 ° and $\theta_{max}$ = 35 °, of which 1184 were independent and 1155 reflections with $I > 2\sigma(I)$. The crystal structure was solved using direct method with SHELXS (Sheldrick, 1997) program and refined via SHELXL (Sheldrick, 2015) program incorporated in the WinGX (Farrugia, 2012) program. Absorption corrections were performed from equivalent reflections on the basis of multiscans.

Magnetic measurements were performed by a Physical Property Measurement System (PPMS) DynaCool magnetometer on selected crystals (7 mg) sealed in a gelatin capsule. The temperature dependence magnetization measurements was recorded between 300 and 1.9 K in both zero field cooled (ZFC) and filed cooled (FC) modes with an applied field ranging from 100 Oe to 40 kOe. The ZFC mode was performed by cooling the sample from 300 down to 2K in the absence of a magnetic field. Afterward, the desired magnetic field was applied and the data were collected on heating the sample up to 300 K. After reaching 300 K, the data were then recorded with the same strength of the field on cooling the sample down to 2 K, i.e., FC. The isothermal magnetization curves were measured up to 90 kOe.

## 2.1. Synthesis and crystallization

Single crystals of this cobalt hydrogen vanadate were grown by hydro-thermal method during the investigation of $Na_2O$—CoO—$V_2O_5$ ternary system. A mixture of $NaNO_3$ ( ≥ 99% , Sigma-Aldrich), $Co(NO_3)_2.6H_2O$ ( ≥ 98.0 %, Scharlau) and $NH_4VO_3$ ( ≥ 98.0 %, Acros Organics), in a molar ratio of Na : Co : V = 7 : 1: 4, were placed in Teflon lined autoclave with 12 ml of distilled water, then transferred to an oven and heated under autogenous pressure at 473 K for 12 days. After cooling-down the autoclave to room temperature, the mixture was washed with distilled water and dried at room temperature. Black hexagonal prismatic crystals were obtained with suitable sizes for X-ray diffraction measurements.

## 2.2. Refinement

Crystal data, data collection and structure refinement details of $Co_{13.5}(OH)_6(H_{0.5}VO_{3.5})_2(VO_4)_6$ are summarized in Table 1. The highest peak and the deepest hole in the final Fourier map are at 0.37 Å from Co2 and 0.49 Å from V1. The not significant inter-atomic bonds and angles were removed from the CIF file.

In the refinement procedure, the vanadium sites V2A and V2B which both occupy the special position 2*b* are constrained to have the same anisotropic displacement parameters, also the occupancy of these sites was refined and restrained to be equal to the occupancy of one site.

On the other hand, the atom O5, attached to V2A and V2B atoms, showed high atomic displacement parameter, consequently, we have allowed the refinement of its occupancy which drops to half occupancy of the site and then fixed, accordingly the occupancy of its riding hydrogen H5 was fixed to half occupancy of the site.

# 3. Results and discussion

## 3.1. Scanning electron microscopy analysis

The morphology and elemental analysis of the obtained single crystals were characterized by scanning electron microscope (SEM) and energy dispersive X-ray spectrometer (EDS). As shown in Figure 1, the SEM micrographs reveal the formation of hexagonal prismatic single crystals with deferent sizes. The EDS spectrum confirmed the existence of only Co, V and oxygen atoms. The Co/V % atomic ratio of 1.65 is in good agreement with the elemental composition of $Co_{13.5}(OH)_6(H_{0.5}VO_{3.5})_2(VO_4)_6$.

## 3.2. Crystal structure

The novel cobalt hydrogen vanadate, $Co_{13.5}(OH)_6(H_{0.5}VO_{3.5})_2(VO_4)_6$ is isostructural to $Mg_{13.4}(OH)_6(HVO_4)_2(H_{0.2}VO_4)_6$ (Đorđević *et al.*, 2008), and to the mineral phospho-ellenbergerite $Mg_{12}(Mg, Fe, \square)_2(PO_4, PO_3OH, AsO_4)_6(PO_3OH, CO_3)_2(OH)_6$ (Raade *et al.*, 1998). The phospho-ellenbergerite mineral and other related structures (Che *et al.*, 2005; Gu *et al.*, 2007; Ni *et al.*, 2009; Amorós *et al.*, 1996; Mentre *et al.*, 2018) were reported as dumortierite-like materials (Evans & Groat, 2012) due to their similar topology to the mineral dumortierite, i.e., $(Al, \square)Al_6(BO_3)Si_3O_{13}(O,OH)_2$. This vanadate crystallizes in hexagonal system with non-centrosymmetric space group, i.e, $P6_3mc$. The asymmetric unit of this vanadate contains two atoms Co1 and O1 in general position 12*d*, Co2 in the special position 2*a (3m)*, V2 and O5 in the special position 2*b (3m)* and the remaining atoms are located on the mirror in the special Wyckoff position 6c (*m*). In fact, the Co2 site is partially occupied with an occupancy rate of 75%. In the isostructural magnesium hydrogen vanadate, i.e, $Mg_{13.4}(OH)_6(HVO_4)_2(H_{0.2}VO_4)_6$ (Đorđević *et al.*, 2008) an occupancy rate of 69.8 % is reported in the similar site. Moreover, this structure is characterized by a disorder in the vanadium site V2 located in the special position 2*b* which is splitted in V2A and V2B positions with the occupancy rates of 61.0 % and 39.0 %, respectively. The refinement of this model led to the exact chemical formula of $Co_{13.5}(OH)_6(H_{0.5}VO_{3.5})_2(VO_4)_6$. Crystal data, data collection and structure refinement details are summarized in Table 1. Fractional atomic coordinates, atomic displacement parameters, geometric parameters are given in supporting information. Hydrogen-bond geometry is presented in table 2.

In this new vanadate, each cobalt atoms is surrounded by six oxygen atoms forming an octahedral environment. The $Co1O_6$ exhibits hydrogen atom attached to O6 forming $Co1O_5(OH)$ octahedra. Two of these octahedra share a common face to form dimers $Co(1)_2O_8(OH)$. The connection of these dimers by sharing edges leads to tetra-mers $Co(1)_4O_{14}(OH)_2$ giving a quadruply bridging hydroxide groups O6H6 (Fig. 2a) . In the tetrameric units, the distance between Co1—Co1 ions in the face-sharing dimer is 2.854 (1) Å and in the edge-sharing one is 3.040 (1) Å. Each of these tetrameric units is surrounded by four other same units via corner-sharing leading to a layer parallel to (110) plane (Fig. 2b). The stacking of these layers along [001] direction yield to 3D framework comprising two type of channels, i.e., hexagonal broader channels and triangular smaller ones as shown in Fig. 2b. The $Co2O_6$ octahedra share a common face to form 1D chains running along the [001] direction. These chains are located in the center of the hexagonal channels and surrounded by $V1O_4$ tetrahedra through vertices sharing (Fig. 3). The distance between Co2—Co2 in these chains is 2.532 (2) Å, which is much shorter than Co1—Co1 in the dimers, which explains the cationic deficit of this site that contains 1/4 of lacuna. As a matter of fact, this deficit is balanced exactly by the hydrogen H5 which is located in the Wyckoff position 6*c* but with an occupancy rate of 0.08333. In this context, it is interesting to compare the present structure with that of $Mg_{13.4}(OH)_6(HVO_4)_2(H_{0.2}VO_4)_6$ in which the principal difference is the electroneutrality obtained by addition of 1.2 Hydrogen atom. The second smaller channels contain $V2O_4$ tetrahedra which are splited in two tetrahedra with different orientations: one oriented down and the other up. The duplicated V2A and V2B atoms exhibit statistical disorder which are both located in the special position 2*b* on the threefold axis as shown in Fig. 4. Both V2A and V2B atoms are four-coordinated and share a common three O4 atoms leading to the formation of trigonal bipyramid with common basal plane, V2A$(O4)_3$O5 and V2B$(O4)_3$O5. The repetition of these trigonal bipyramids along the c-axis by sharing O5 atom creates infinite chains (Fig. 5). Also, this new vanadate is characterized by the presence of bifurcated hydrogen bonds as shown in Table 2 and Fig. 5.

### 3.3. Magnetic properties

The magnetic susceptibility ($\chi = M/H$) and the inverse magnetic susceptibility $\chi^{-1}$ for $Co_{13.5}(OH)_6(H_{0.5}VO_{3.5})_2(VO_4)_6$, measured at 10 Oe in ZFC mode within the temperature range of 1.9-300 K are presented in Fig. 6. The data were corrected from diamagnetic core ($-6.5204.10^{-4}$ emu.mol$^{-1}$). In ZFC mode, during heating the sample, a sharp peak at 9 K is observed and the susceptibility drops to almost zero for further heating up to 300 K. The high temperature region above 70 K of $1/\chi$ versus $T$ follows well the Curie-Weiss law. The fitting gave a paramagnetic Curie temperature $\theta_{cw}$ = - 47.13 K indicating predominant antiferromagnetic interactions between $Co^{2+}$ cations. The obtained Curie constant $C$ = 3.37 emu K mol$^{-1}$ per $Co^{2+}$, yielded to an effective magnetic moment per $Co^{2+}$ of $\mu_{eff}$ = 5.19 $\mu_B$. This value is in good agreement with the calculated considering spin-orbit coupling hypothesis for $Co^{2+}$ ($d^7$, $S$ = 3/2, $\mu_{LS}$ = 5.20 $\mu_B$) due to the unquenched orbital contribution commonly observed in $Co^{2+}$, indicating strong spin–orbital coupling and anisotropy for $Co^{2+}$ (Viola *et al.*, 2003; Nakayama *et al.*, 2013; He *et al.*, 2005).

Figure 7 shows the magnetic susceptibility versus $T$ in both field-cooled (FC) and zero-field-cooled (ZFC) modes at different applied fields in low-temperature region of 1.9-40 K. We observe at 100 Oe in both ZFC and FC a sharp peak and a strong divergence between ZFC and FC curves at low temperature, at 500 and 1000 Oe, the peak became broaden, reduces the height and shifts towards lower temperature with maintaining the bifurcation. At the field of 5 kOe the peak is almost suppressed and the bifurcation became nearly negligible. For fields higher than 5 kOe the transition and the bifurcation were completely vanished. All these facts are a clear signal of weak ferromagnetic correlation or Spin-Glass type behavior (Belik *et al.*, 2007; Edwards & Anderson, 1975; Goldfarb *et al.*, 1985).

The origin of this weak ferromagnetic correlation might be related to site disorder and geometrical frustration which both leads to frustrated magnetic interactions (Ramirez, 1994). The title compound is characterized by a disorder in Co2 site. Furthermore, an empirical measure of frustration by calculating the quantity defined by (Ramirez, 1994) : $f = -\theta_{cw}/T_N$, where $\theta_{cw}$ the Weiss temperature and $T_N$ is the ordering temperature. A value of $f > 1$ indicates the frustration occurs in the system. The calculated value of $f$ = 5.24 considering the Weiss temperature of $\theta_{cw}$ = - 47.13 K and the ordering temperature of $T_N$ = 9 K, clearly designates the frustration in the system.

Figure 8, represent the magnetization as function of the field measured at 2.5 and 10 K. As shown in Fig. 8a, the magnetization at 2.5 and 10 K increases as increasing the field without saturation even at 90 kOe reaching a values of ~1 $\mu_B$ at 2.5 K and ~0.95 $\mu_B$ at 10 K. These values are approximately 1/3 of saturation moment of $Co^{2+}$ ( 3 $\mu_B$, $S$ = 3/2). The observed narrow hysteresis loop (inset of Fig. 8b) confirms the week ferromagnetic component resulting from the frustrated magnetic interactions.

## 4. Conclusion

In this work we have successfully synthesized the single crystals of a new geometrically frustrated cobalt hydrogen vanadate displaying dumortierite-like structure, i.e., $Co_{13.5}(OH)_6(H_{0.5}VO_{3.5})_2(VO_4)_6$. The single crystals of this new magnetic material were characterized by single crystal X-ray diffraction and scanning electron microscopy. The structure of this new material is built-up by the succession of $CoO_6$ octahedra and $VO_4$ tetrahedra, which yields to 3D framework comprising two type of channels, i.e., hexagonal channels containing 1D cobalt-chains and triangular channels. The magnetic investigation revealed antiferromagnetic behavior with the existence of weak ferromagnetic correlation. However, deep magnetic investigations are needed to shed light on the magnetic behavior of this vanadate

***Table 1***

*Experimental details*

| | |
|---|---|
| Crystal data | |
| Chemical formula | $H_7Co_{13.50}O_{37}V_8$ |
| $M_r$ | 1802.13 |
| Crystal system, space group | Hexagonal, $P6_3mc$ |
| Temperature (K) | 293 |
| $a$, $c$ (Å) | 12.837 (8), 5.064 (3) |
| $V$ (Å$^3$) | 722.8 (10) |
| $Z$ | 1 |
| Radiation type | Mo $K\alpha$ |
| $\mu$ (mm$^{-1}$) | 10.09 |
| Crystal size (mm) | 0.30 × 0.20 × 0.19 |
| | |
| Data collection | |
| Diffractometer | Bruker D8 VENTURE Super DUO |
| Absorption correction | Multi-scan (*SADABS*; Krause *et al.*, 2015) |
| $T_{min}$, $T_{max}$ | 0.619, 0.747 |
| No. of measured, independent and observed [$I > 2\sigma(I)$] reflections | 32344, 1184, 1155 |
| $R_{int}$ | 0.041 |
| $(\sin \theta/\lambda)_{max}$ (Å$^{-1}$) | 0.806 |
| | |
| Refinement | |
| $R[F^2 > 2\sigma(F^2)]$, $wR(F^2)$, $S$ | 0.015, 0.043, 1.10 |
| No. of reflections | 1184 |
| No. of parameters | 63 |
| No. of restraints | 1 |
| H-atom treatment | H-atom parameters constrained |
| $\Delta\rangle_{max}$, $\Delta\rangle_{min}$ (e Å$^{-3}$) | 1.21, -0.47 |
| Absolute structure | Flack x determined using 521 quotients [(I+)-(I-)]/[(I+)+(I-)] (Parsons and Flack (2013). |
| Absolute structure parameter | 0.020 (8) |

Computer programs: *APEX3* (Bruker, 2016), *SAINT* (Bruker, 2016), *SAINT*, *SHELXS2014*/7 (Sheldrick, 2015*a*), *SHELXL2014*/7 (Sheldrick, 2015*b*), *DIAMOND*(Brandenburg, 2006), *publCIF* (Westrip, 2010).

### *Table 2*

*Hydrogen-bond geometry (Å, º) for (shelx)*

| *D*—H···*A* | *D*—H | H···*A* | *D*···*A* | *D*—H···*A* |
|---|---|---|---|---|
| O5—H5···O4$^{i}$ | 0.82 | 2.49 | 3.071 (11) | 128 |
| O5—H5···O6$^{ii}$ | 0.82 | 2.35 | 3.141 (3) | 161 |
| O6—H6···O4$^{iii}$ | 0.86 | 2.32 | 3.137 (4) | 159 |
| O6—H6···O5$^{iii}$ | 0.86 | 2.53 | 3.141 (3) | 128 |

Symmetry codes: (i) *x*, *y*, *z*+1; (ii) -*x*+1, -*y*+1, *z*+1/2; (iii) -*x*+1, -*y*+1, *z*-1/2.

### Acknowledgements

This work was done with the support of CNRST (Centre National pour la Recherche Scientifique et Technique) in the Excellence Research Scholarships Program. This work was also supported by the European H2020-MC-RISE-ENGIMA action (n° 778072).

## References

Aasland, S., Fjellvåg, H. & Hauback, B. (1997). *Solid State Communications*. **101**, 187–192.

Aliev, A., Huvé, M., Colis, S., Colmont, M., Dinia, A. & Mentré, O. (2012). *Angewandte Chemie International Edition*. **51**, 9393–9397.

Amorós, P., Marcos, M. D., Roca, M., Beltrán-Porter, A. & Beltrán-Porter, D. (1996). *Journal of Solid State Chemistry*. **126**, 169–176.

Balents, L. (2010). *Nature*. **464**, 199–208.

Belik, A. A., Tsujii, N., Huang, Q., Takayama-Muromachi, E. & Takano, M. (2007). *Journal of Physics: Condensed Matter*. **19**, 145221.

Bramwell, S. T. (2001). *Science*. **294**, 1495–1501.

Che, R. C., Peng, L.-M. & Zhou, W. Z. (2005). *Appl. Phys. Lett.* **87**, 173122.

Đorđević, T., Karanović, Lj. & Tillmanns, E. (2008). *Crystal Research and Technology*. **43**, 1202–1209.

Edwards, S. F. & Anderson, P. W. (1975). *Journal of Physics F: Metal Physics*. **5**, 965–974.

Evans, R. J. & Groat, L. A. (2012). *The Canadian Mineralogist*. **50**, 1197–1231.

Farrugia, L. J. (2012). *Journal of Applied Crystallography*. **45**, 849–854.

Gambardella, P. (2003). *Science*. **300**, 1130–1133.

Goldfarb, R. B., Rao, K. V. & Chen, H. S. (1985). *Solid State Communications*. **54**, 799–801.

Gu, Z., Zhai, T., Gao, B., Zhang, G., Ke, D., Ma, Y. & Yao, J. (2007). *Crystal Growth & Design*. **7**, 825–830.

Hardy, V., Lambert, S., Lees, M. R. & McK. Paul, D. (2003). *Physical Review B*. **68**,.

Harris, M. J., Bramwell, S. T., McMorrow, D. F., Zeiske, T. & Godfrey, K. W. (1997). *Physical Review Letters*. **79**, 2554–2557.

He, Z., Fu, D., Kyômen, T., Taniyama, T. & Itoh, M. (2005). *Chemistry of Materials*. **17**, 2924–2926.

Kageyama, H., Yoshimura, K., Kosuge, K., Azuma, M., Takano, M., Mitamura, H. & Goto, T. (1997). *Journal of the Physical Society of Japan*. **66**, 3996–4000.

Mentre, O., Blazquez-Alcover, I., Garcia-Martin, S., Duttine, M., Wattiaux, A., Simon, P., Huve, M. & Daviero-Minaud, S. (2018). *Inorg. Chem.* **57**, 15093–15104.

Moessner, R. & Ramirez, A. P. (2006). *Physics Today*. **59**, 24–29.

Moulton, B., Lu, J., Hajndl, R., Hariharan, S. & Zaworotko, M. J. (2002). *Angewandte Chemie International Edition*. **41**, 2821–2824.

Nakayama, G., Hara, S., Sato, H., Narumi, Y. & Nojiri, H. (2013). *Journal of Physics: Condensed Matter*. **25**, 116003.

Ni, Y., Liao, K., Hong, J. & Wei, X. (2009). *CrystEngComm*. **11**, 570.

Okamoto, Y., Nohara, M., Aruga-Katori, H. & Takagi, H. (2007). *Physical Review Letters*. **99**,.

Raade, G., Romming, C. & Medenbach, O. (1998). *Mineralogy and Petrology*. **62**, 89–101.

Ramirez, A. P. (1994). *Annual Review of Materials Science*. **24**, 453–480.

Sheldrick, G. M. (1997). SHELXS-97, Program for crystal structure solution University of Göttingen, Germany Göttingen.

Sheldrick, G. M. (2015). *Acta Crystallographica Section C Structural Chemistry*. **71**, 3–8.

Vilminot, S., Baker, P. J., Blundell, S. J., Sugano, T., André, G. & Kurmoo, M. (2010). *Chemistry of Materials*. **22**, 4090–4095.

Viola, M. C., Martínez-Lope, M. J., Alonso, J. A., Martínez, J. L., De Paoli, J. M., Pagola, S., Pedregosa, J. C., Fernández-Díaz, M. T. & Carbonio, R. E. (2003). *Chemistry of Materials*. **15**, 1655–1663.

Wang, R. F., Nisoli, C., Freitas, R. S., Li, J., McConville, W., Cooley, B. J., Lund, M. S., Samarth, N., Leighton, C., Crespi, V. H. & Schiffer, P. (2006). *Nature*. **439**, 303–306.

Zhang, X.-M., Li, C.-R., Zhang, X.-H., Zhang, W.-X. & Chen, X.-M. (2008). *Chemistry of Materials*. **20**, 2298–2305.

Zheng, Y.-Z., Tong, M.-L., Xue, W., Zhang, W.-X., Chen, X.-M., Grandjean, F. & Long, G. J. (2007). *Angewandte Chemie International Edition*. **46**, 6076–6080.

Westrip, S. P. (2010). *J. Appl. Cryst.* **43**, 920–925.

**Figure 1**

SEM micrographs and EDS spectrum showing the morphology and the elemental composition of the synthesized single crystals

**Figure 2**

(a) Tetrameric units in the structure of $Co_{13.5}(OH)_6(H_{0.5}VO_{3.5})_2(VO_4)_6$. (b) Layer parallel to (110) plane

**Figure 3**

(a) Chain parallel to the *c*-axis formed by face-sharing $Co2O_6$ octahedra surrounded by $V1O_4$ tetrahedra, (b) view perpendicular to (110) plane, (c) view in the hexagonal channels

**Figure 4**

Ortep plot showing V2A and V2B atoms located on the threefold axis

**Figure 5**

Polyhedral representation showing trigonal bipyramids located in the triangular channels running along the *c*-axis

**Figure 6**

Plots of susceptibility and reciprocal susceptibility. The inset figure represents the Curie-Weiss fit of 1/ $\chi$ in the high temperature paramagnetic phase.

**Figure 7**

Temperature dependence of magnetic susceptibility at different applied fields showing a clear divergence between the ZFC and FC curves under 5 kOe.

**Figure 8**

(a) *M-H* curves at 2.5 and 10 K measured up to 90 kOe. (b) Hysteresis loop measured at 2.5 K between -90 kOe and 90 kOe

**supporting information**

# Magnetic Properties of a New Cobalt Hydrogen Vanadate with Dumortierite-Like Structure: $Co_{13.5}(OH)_6(H_{0.5}VO_{3.5})_2(VO_4)_6$

**Mohammed Hadouchi,* Abderrazzak Assani, Mohamed Saadi, Abdelilah Lahmar, Mimoun El Marssi and Lahcen El Ammari**

## Computing details

Data collection: *APEX3* (Bruker, 2016); cell refinement: *SAINT* (Bruker, 2016); data reduction: *SAINT*; program(s) used to solve structure: *SHELXS2014*/7 (Sheldrick, 2015*a*); program(s) used to refine structure: *SHELXL2014*/7 (Sheldrick, 2015*b*); molecular graphics: *DIAMOND*(Brandenburg, 2006); software used to prepare material for publication: *publCIF* (Westrip, 2010).

## (shelx)

### *Crystal data*

| | |
|---|---|
| $H_7Co_{13.50}O_{37}V_8$ | $D_x$ = 4.140 Mg m$^{-3}$ |
| $M_r$ = 1802.13 | Mo *K*α radiation, λ = 0.71073 Å |
| Hexagonal, *P*6$_3$*mc* | Cell parameters from 1184 reflections |
| *a* = 12.837 (8) Å | θ = 3.2–35.0° |
| *c* = 5.064 (3) Å | μ = 10.09 mm$^{-1}$ |
| *V* = 722.8 (10) Å$^3$ | *T* = 293 K |
| *Z* = 1 | Hexagonal prism, black |
| *F*(000) = 852 | 0.30 × 0.20 × 0.19 mm |

### *Data collection*

| | |
|---|---|
| Bruker D8 VENTURE Super DUO diffractometer | 1184 independent reflections |
| Radiation source: INCOATEC IμS micro-focus source | 1155 reflections with *I* > 2σ(*I*) |
| Detector resolution: 10.4167 pixels mm$^{-1}$ | $R_{int}$ = 0.041 |
| φ and ω scans | $θ_{max}$ = 35.0°, $θ_{min}$ = 3.2° |
| Absorption correction: multi-scan (*SADABS*; Krause *et al.*, 2015) | *h* = -20→20 |
| $T_{min}$ = 0.619, $T_{max}$ = 0.747 | *k* = -20→20 |
| 32344 measured reflections | *l* = -8→8 |

## *Refinement*

| | |
|---|---|
| Refinement on $F^2$ | H-atom parameters constrained |
| Least-squares matrix: full | $w = 1/[\sigma^2(F_o^2) + (0.0211P)^2 + 0.7866P]$ where $P = (F_o^2 + 2F_c^2)/3$ |
| $R[F^2 > 2\sigma(F^2)] = 0.015$ | $(\Delta/\sigma)_{max} = 0.001$ |
| $wR(F^2) = 0.043$ | $\Delta\rangle_{max} = 1.21$ e Å$^{-3}$ |
| $S = 1.10$ | $\Delta\rangle_{min} = -0.47$ e Å$^{-3}$ |
| 1184 reflections | Extinction correction: *SHELXL*, $Fc^*=kFc[1+0.001xFc^2\lambda^3/\sin(2\theta)]^{-1/4}$ |
| 63 parameters | Extinction coefficient: 0.0056 (6) |
| 1 restraint | Absolute structure: Flack x determined using 521 quotients [(I+)-(I-)]/[(I+)+(I-)] (Parsons and Flack (2013). |
| Hydrogen site location: mixed | Absolute structure parameter: 0.020 (8) |

## *Special details*

| |
|---|
| *Geometry*. All esds (except the esd in the dihedral angle between two l.s. planes) are estimated using the full covariance matrix. The cell esds are taken into account individually in the estimation of esds in distances, angles and torsion angles; correlations between esds in cell parameters are only used when they are defined by crystal symmetry. An approximate (isotropic) treatment of cell esds is used for estimating esds involving l.s. planes. |

## *Fractional atomic coordinates and isotropic or equivalent isotropic displacement parameters (Å$^2$) for (shelx)*

| | *x* | *y* | *z* | $U_{iso}$*/$U_{eq}$ | Occ. (<1) |
|---|---|---|---|---|---|
| V1 | 0.69852 (5) | 0.84926 (2) | 0.62349 (10) | 0.00656 (10) | |
| V2A | 0.3333 | 0.6667 | 0.3690 (4) | 0.0070 (4) | 0.610 (5) |
| V2B | 0.3333 | 0.6667 | 0.0920 (6) | 0.0061 (5) | 0.390 (5) |
| Co1 | 0.57335 (2) | 0.64904 (3) | 0.09342 (5) | 0.00765 (8) | |
| Co2 | 1.0000 | 1.0000 | 0.9265 (3) | 0.0127 (2) | 0.75 |
| O1 | 0.65816 (17) | 0.72332 (15) | 0.4391 (4) | 0.0098 (3) | |
| O2 | 0.6119 (2) | 0.80595 (12) | 0.9087 (5) | 0.0099 (4) | |
| O3 | 0.8507 (2) | 0.92536 (12) | 0.6824 (7) | 0.0145 (5) | |
| O4 | 0.40426 (12) | 0.59574 (12) | 0.2403 (7) | 0.0155 (5) | |
| O5 | 0.3333 | 0.6667 | 0.720 (2) | 0.0174 (19) | 0.5 |
| H5 | 0.3681 | 0.6319 | 0.7739 | 0.026* | 0.1667 |
| O6 | 0.52603 (12) | 0.47397 (12) | 0.2789 (5) | 0.0090 (4) | |
| H6 | 0.5536 | 0.4464 | 0.1584 | 0.013* | |

*Atomic displacement parameters (Å$^2$) for (shelx)*

| | $U^{11}$ | $U^{22}$ | $U^{33}$ | $U^{12}$ | $U^{13}$ | $U^{23}$ |
|---|---|---|---|---|---|---|
| V1 | 0.0088 (2) | 0.00597 (14) | 0.0059 (2) | 0.00438 (10) | -0.00049 (18) | -0.00024 (9) |
| V2A | 0.0040 (4) | 0.0040 (4) | 0.0131 (7) | 0.0020 (2) | 0.000 | 0.000 |
| V2B | 0.0057 (6) | 0.0057 (6) | 0.0068 (9) | 0.0029 (3) | 0.000 | 0.000 |
| Co1 | 0.00715 (13) | 0.00836 (13) | 0.00721 (13) | 0.00369 (10) | 0.00000 (11) | -0.00030 (13) |
| Co2 | 0.0041 (2) | 0.0041 (2) | 0.0298 (6) | 0.00207 (11) | 0.000 | 0.000 |
| O1 | 0.0136 (7) | 0.0082 (6) | 0.0085 (7) | 0.0060 (6) | -0.0016 (6) | -0.0005 (6) |
| O2 | 0.0140 (11) | 0.0093 (7) | 0.0080 (10) | 0.0070 (5) | 0.0024 (9) | 0.0012 (4) |
| O3 | 0.0082 (10) | 0.0132 (8) | 0.0203 (13) | 0.0041 (5) | -0.0014 (9) | -0.0007 (5) |
| O4 | 0.0092 (8) | 0.0092 (8) | 0.0293 (15) | 0.0055 (9) | 0.0016 (5) | -0.0016 (5) |
| O5 | 0.015 (3) | 0.015 (3) | 0.023 (5) | 0.0073 (13) | 0.000 | 0.000 |
| O6 | 0.0093 (7) | 0.0093 (7) | 0.0086 (9) | 0.0048 (9) | 0.0009 (4) | -0.0009 (4) |

*Geometric parameters (Å, º) for (shelx)*

| | | | |
|---|---|---|---|
| V1—O1 | 1.7079 (18) | Co1—O6 | 2.222 (2) |
| V1—O1$^{i}$ | 1.7080 (18) | Co2—O3$^{vi}$ | 2.069 (3) |
| V1—O3 | 1.718 (3) | Co2—O3 | 2.069 (3) |
| V1—O2 | 1.736 (3) | Co2—O3$^{vii}$ | 2.070 (3) |
| V2A—O4 | 1.706 (3) | Co2—O3$^{viii}$ | 2.106 (3) |
| V2A—O4$^{ii}$ | 1.706 (3) | Co2—O3$^{ix}$ | 2.106 (3) |
| V2A—O4$^{i}$ | 1.706 (3) | Co2—O3$^{x}$ | 2.106 (3) |
| V2A—O5 | 1.777 (13) | O1—Co1$^{ix}$ | 2.0262 (17) |
| V2B—O4 | 1.747 (3) | O2—Co1$^{xi}$ | 2.0446 (13) |
| V2B—O4$^{i}$ | 1.747 (3) | O2—Co1$^{xii}$ | 2.0446 (13) |
| V2B—O4$^{ii}$ | 1.747 (3) | O3—Co2$^{xiii}$ | 2.106 (3) |
| V2B—O5$^{iii}$ | 1.884 (13) | O4—Co1$^{xiv}$ | 2.061 (2) |
| Co1—O1$^{iv}$ | 2.026 (2) | O5—H5 | 0.8200 |
| Co1—O1 | 2.030 (2) | O6—Co1$^{ix}$ | 2.155 (2) |
| Co1—O2$^{iii}$ | 2.0447 (13) | O6—Co1$^{xv}$ | 2.155 (2) |
| Co1—O4 | 2.061 (2) | O6—Co1$^{xiv}$ | 2.222 (2) |
| Co1—O6$^{v}$ | 2.155 (2) | O6—H6 | 0.8643 |
| | | | |
| O1—V1—O1$^{i}$ | 105.30 (13) | O1$^{iv}$—Co1—O6$^{v}$ | 87.39 (8) |
| O1—V1—O3 | 110.72 (9) | O1—Co1—O6$^{v}$ | 164.50 (8) |
| O1$^{i}$—V1—O3 | 110.72 (10) | O2$^{iii}$—Co1—O6$^{v}$ | 98.44 (9) |
| O1—V1—O2 | 108.01 (8) | O4—Co1—O6$^{v}$ | 83.34 (10) |
| O1$^{i}$—V1—O2 | 108.01 (9) | O1$^{iv}$—Co1—O6 | 89.59 (8) |
| O3—V1—O2 | 113.69 (15) | O1—Co1—O6 | 85.52 (8) |
| O4—V2A—O4$^{ii}$ | 106.33 (15) | O2$^{iii}$—Co1—O6 | 177.40 (9) |
| O4—V2A—O4$^{i}$ | 106.33 (15) | O4—Co1—O6 | 78.57 (9) |
| O4$^{ii}$—V2A—O4$^{i}$ | 106.33 (14) | O6$^{v}$—Co1—O6 | 78.99 (5) |
| O4—V2A—O5 | 112.45 (13) | O3$^{vi}$—Co2—O3 | 87.98 (14) |
| O4$^{ii}$—V2A—O5 | 112.45 (13) | O3$^{vi}$—Co2—O3$^{vii}$ | 87.98 (14) |
| O4$^{i}$—V2A—O5 | 112.45 (13) | O3—Co2—O3$^{vii}$ | 87.98 (14) |
| O4—V2B—O4$^{i}$ | 102.87 (17) | O3$^{vi}$—Co2—O3$^{viii}$ | 92.95 (7) |
| O4—V2B—O4$^{ii}$ | 102.87 (17) | O3—Co2—O3$^{viii}$ | 178.70 (16) |
| O4$^{i}$—V2B—O4$^{ii}$ | 102.87 (17) | O3$^{vii}$—Co2—O3$^{viii}$ | 92.96 (8) |
| O4—V2B—O5$^{iii}$ | 115.46 (14) | O3$^{vi}$—Co2—O3$^{ix}$ | 178.70 (16) |
| O4$^{i}$—V2B—O5$^{iii}$ | 115.46 (14) | O3—Co2—O3$^{ix}$ | 92.96 (7) |
| O4$^{ii}$—V2B—O5$^{iii}$ | 115.46 (14) | O3$^{vii}$—Co2—O3$^{ix}$ | 92.96 (7) |
| O1$^{iv}$—Co1—O1 | 91.98 (8) | O3$^{viii}$—Co2—O3$^{ix}$ | 86.09 (14) |
| O1$^{iv}$—Co1—O2$^{iii}$ | 90.67 (9) | O3$^{vi}$—Co2—O3$^{x}$ | 92.96 (7) |
| O1—Co1—O2$^{iii}$ | 97.06 (9) | O3—Co2—O3$^{x}$ | 92.95 (8) |
| O1$^{iv}$—Co1—O4 | 166.15 (8) | O3$^{vii}$—Co2—O3$^{x}$ | 178.70 (16) |
| O1—Co1—O4 | 94.20 (11) | O3$^{viii}$—Co2—O3$^{x}$ | 86.09 (14) |
| O2$^{iii}$—Co1—O4 | 100.84 (10) | O3$^{ix}$—Co2—O3$^{x}$ | 86.09 (14) |

Symmetry codes: (i) *x*, *x*-*y*+1, *z*; (ii) -*x*+*y*, *y*, *z*; (iii) *x*, *y*, *z*-1; (iv) *y*, *x*, *z*-1/2; (v) -*x*+1, -*y*+1, *z*-1/2; (vi) -*y*+2, -*x*+2, *z*; (vii) -*x*+*y*+1, *y*, *z*; (viii) -*x*+2, -*y*+2, *z*+1/2; (ix) *y*, *x*, *z*+1/2; (x) *x*-*y*+1, -*y*+2, *z*+1/2; (xi) *x*, *x*-*y*+1, *z*+1; (xii) *x*, *y*, *z*+1; (xiii) -*x*+2, -*y*+2, *z*-1/2; (xiv) -*y*+1, -*x*+1, *z*; (xv) -*x*+1, -*y*+1, *z*+1/2.

### *Hydrogen-bond geometry (Å, º) for (shelx)*

| *D*—H···*A* | *D*—H | H···*A* | *D*···*A* | *D*—H···*A* |
|---|---|---|---|---|
| O5—H5···O4$^{xii}$ | 0.82 | 2.49 | 3.071 (11) | 128 |
| O5—H5···O6$^{xv}$ | 0.82 | 2.35 | 3.141 (3) | 161 |
| O6—H6···O4$^{v}$ | 0.86 | 2.32 | 3.137 (4) | 159 |
| O6—H6···O5$^{v}$ | 0.86 | 2.53 | 3.141 (3) | 128 |

Symmetry codes: (v) -*x*+1, -*y*+1, *z*-1/2; (xii) *x*, *y*, *z*+1; (xv) -*x*+1, -*y*+1, *z*+1/2.